\documentclass[prl,twocolumn,showpacs,superscriptaddress,floatfix]{revtex4}
\usepackage{graphicx}%
\begin{document}

\title{Force Chains, Microelasticity and Macroelasticity} 
\author{C. Goldenberg} 
\email{chayg@post.tau.ac.il}
\affiliation{School of Physics and Astronomy,
 Tel-Aviv University, Ramat-Aviv, Tel-Aviv 69978, Israel}
\author{I. Goldhirsch} 
\email{isaac@eng.tau.ac.il}
\affiliation{Fluid Mechanics and Heat Transfer,
 Tel-Aviv University, Ramat-Aviv, Tel-Aviv 69978, Israel}
\date{December 28, 2001}
\begin{abstract}
  It has been claimed that quasistatic granular materials, as well as nanoscale
  materials, exhibit departures from elasticity even at small loadings. It is
  demonstrated, using 2D and 3D models with interparticle harmonic
  interactions, that such departures are expected at small scales [below
  ${\cal{O}}(100)$ particle diameters], at which continuum elasticity is
  invalid, and vanish at large scales.  The models exhibit force chains on
  small scales, and force and stress distributions which agree with
  experimental findings.  Effects of anisotropy, disorder and boundary
  conditions are discussed as well.
\end{abstract}

\maketitle

There are at least two classes of systems whose apparent departure from elastic
mechanical response, even for infinitesimal loads, has been recently discussed
in the literature: granular~\cite{ChaosGranularFocus} and nanoscale~\cite{Nano}
materials. These two classes share a common property: both are typically
composed of a relatively small number of constituents, macroscopic grains in
the former and {\it atoms} in the latter case. Nanoscale grains are not
considered here.  The discussion below, though relevant to both classes, is
worded in ``granular terms'' and focuses on granular materials.

Models whereby forces in static granular matter ``propagate'' (corresponding to
hyperbolic or parabolic PDE's) have been suggested~\cite{Cates99,ForceChains}.
This mechanism is in marked contrast with the non-propagating nature of the
classical equations of static elasticity (elasto-plastic models are
successfully used by engineers in this field ~\cite{Nedderman92,Savage97}).

One of the goals of this Letter is to show that some recent
experiments~\cite{Geng00,Clement}, in which the quasi-static response of
granular matter has been measured, are consistent with an elastic description,
up to differences which stem from the fact that continuum elasticity is a
macroscopic theory, valid only above a certain spatial scale.  A crossover
between microelasticity, i.e., the small scale response of a system whose
constituents interact by harmonic forces, and continuum elasticity
(`macroelasticity') at large scales is demonstrated.  Note that both the
response of small (elastic) systems and the short distance response of {\it
  any} such system are microelastic.

Any particle in a granular medium experiences interactions with a finite number
of other particles, each of which defines a different direction.  Therefore the
local environment of a particle is not isotropic.  The existence of preferred
directions on the particle scale implies the possible emergence of force
chains, i.e., chains of contacts along which the forces are e.g., stronger than
the mean interparticle force.  This fact does not preclude elasticity, e.g.,
the forces may be derived from a potential (and can be linearized) yielding a
set of discrete elliptic equations. The notion of force propagation along force
chains is merely an \textit{approximation} pertaining to the ``strong'' forces.
Continuum theory cannot be expected to describe these microscopic interparticle
\textit{forces}.  The macroscopic field which is related to the microscopic
forces is the \textit{stress}.  Even at small but finite spatial resolution,
the stress tensor is determined by an appropriate average over forces, and it
may bear little or no resemblance to the force distribution on the microscopic
(particle) scale. Though the stress field is well defined on small scales, the
constitutive relations correspond to continuum elasticity only on sufficiently
large scales. In particular, disorder can have a large effect at small scales,
yet allow for large scale homogeneous (and possibly isotropic) elasticity.

The discussion below does not apply  to isostatic systems or systems at
incipient failure, where particle rearrangements are prominent
and the range of validity of  elasticity  may be extremely limited.
We  consider only dense systems (which form the majority of granular systems in
practical applications) where geometrical constraints and friction prevent
major rearrangements for sufficiently small loads  (depending on the yield
stress).

\begin{figure}[!ht]
\includegraphics[width=3.25in]{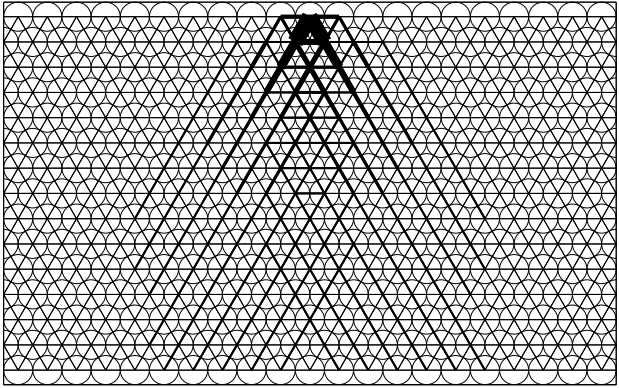}
\caption{Force chains in a 2D triangular lattice. A unit vertical force is
  applied to the center particle in the top layer, with no gravity. Line widths
  for all the lines shown are proportional to the forces. The central region of
  the lattice is shown (the entire lattice comprises 15 layers of 41 particles
  each).}
\label{fig:lat2Dhex_particles+forces}
\end{figure}

\begin{figure}[!ht]
\begin{center}
  \includegraphics[width=3.25in]{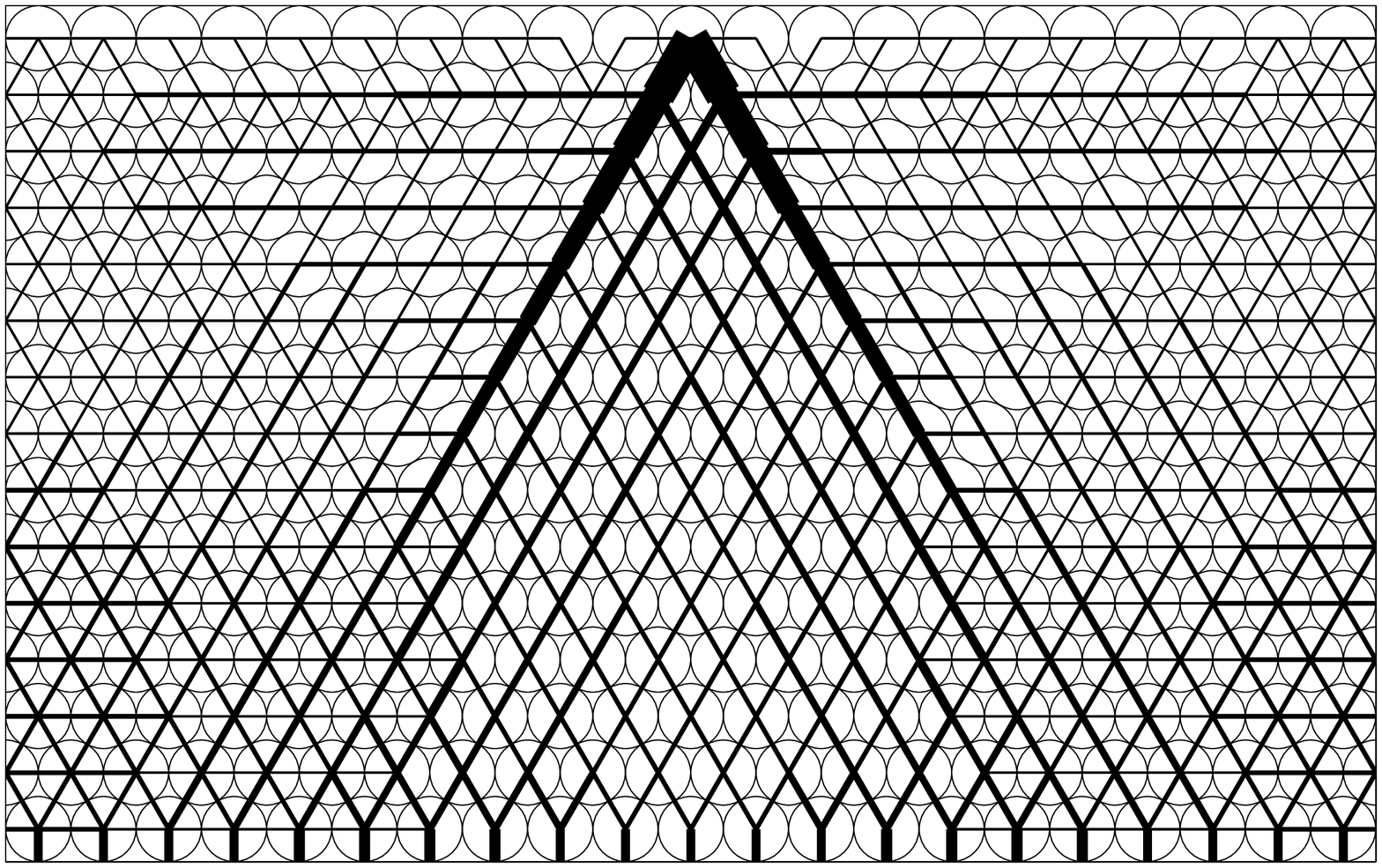}
\end{center}
\caption{Force chains in the same 2D triangular lattice as in
  Fig.~\ref{fig:lat2Dhex_particles+forces}, but with one-sided springs and 
  gravity.}
\label{fig:lat2Dhexnotens_particles+forces}
\end{figure}

\begin{figure}[!ht]
\includegraphics[width=3.25in]{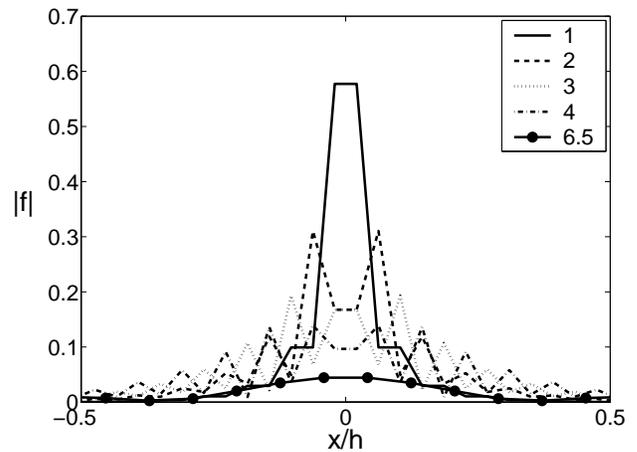}
\caption{The norms of the interparticle forces, $|\mathbf{f}|$, in the system
  depicted in Fig.~\ref{fig:lat2Dhex_particles+forces}, vs. the horizontal
  position, $x$, at several depths. The legend indicates the depth measured
  from the top, in layer numbers.}
\label{fig:forcesdiffdepth}
\end{figure}

\begin{figure}[!ht]
\includegraphics[width=3.25in]{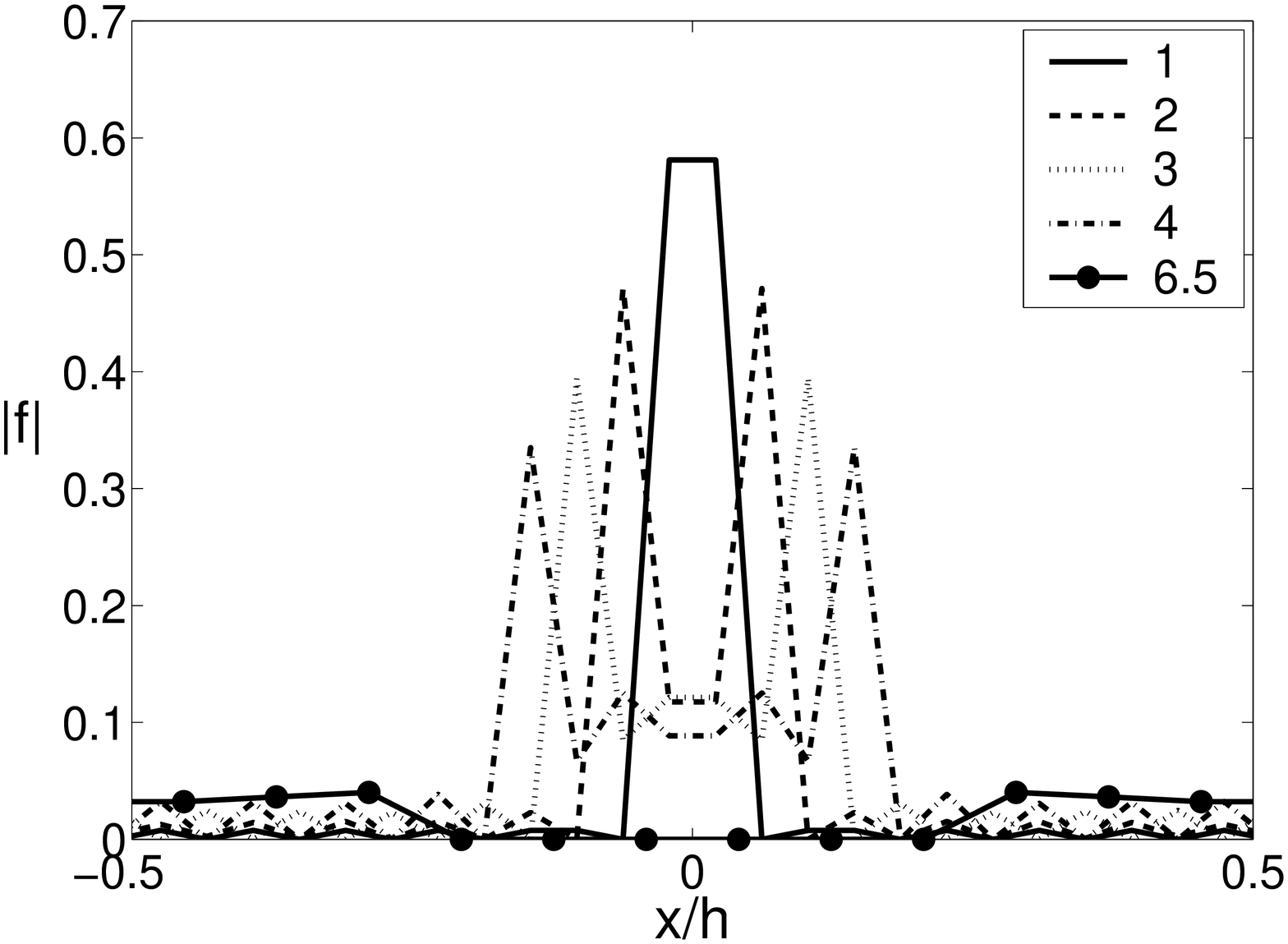}
\caption{Same as Fig.~\ref{fig:forcesdiffdepth}, for the case of one-sided
  springs.}
\label{fig:notensionforcesdiffdepth}
\end{figure}

\begin{figure}[!ht]
  \includegraphics[width=3.25in]{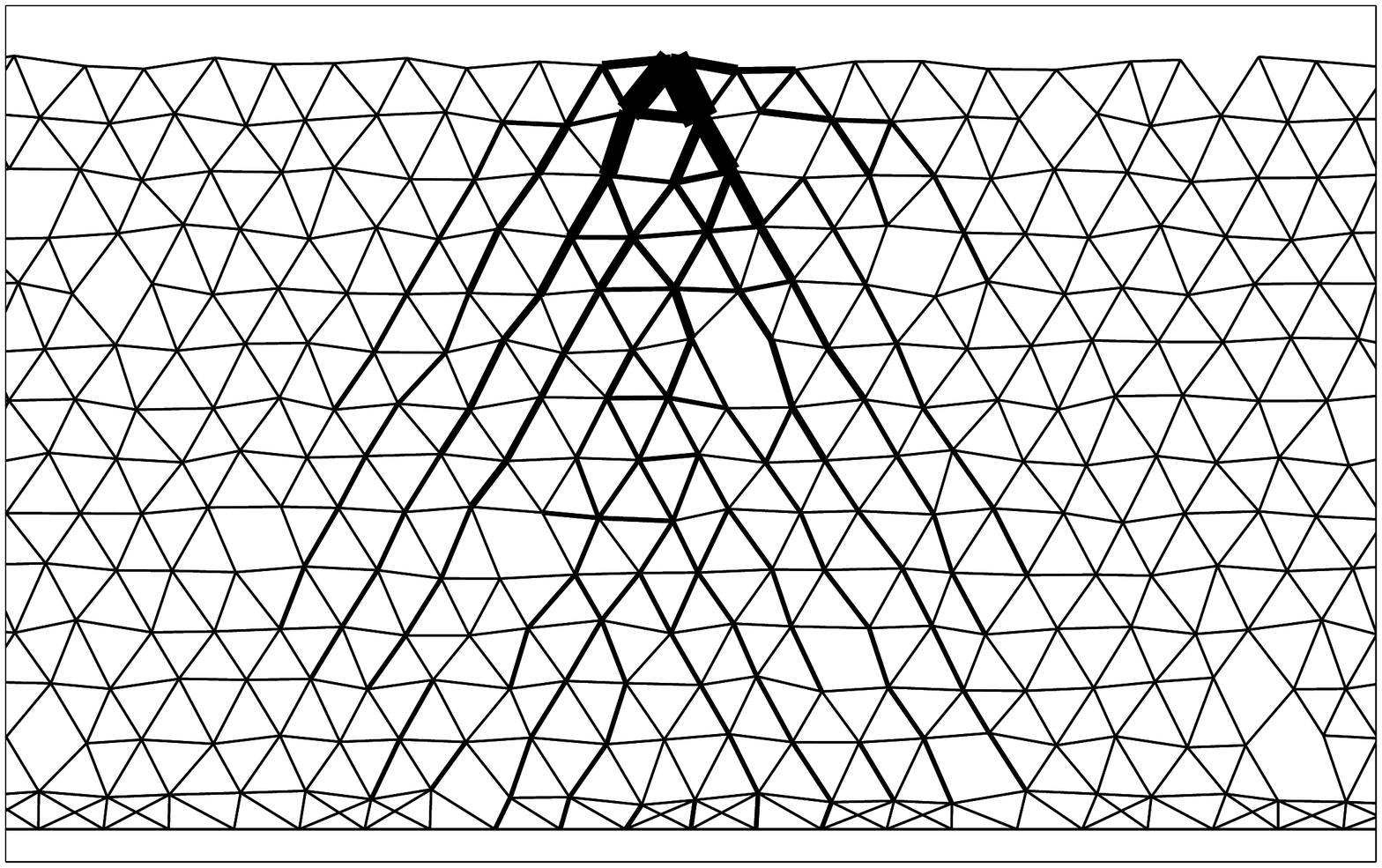}
\caption{Force chains in a random network of springs with
  $c_\mathrm{max}=1.2d$ (see text).}
\label{fig:pertlowconlat2Dhex_particles+forces}
\end{figure}

Consider a two dimensional (2D) collection of uniform disks resting in a finite
enclosure (with a horizontal `floor') under the influence of gravity. A unit
downward vertical force is applied at the top of the system (in the simulations
below the force  acts  in the middle of the upper layer).
Fig.~\ref{fig:lat2Dhex_particles+forces} presents the results of a
simulation in which the disks are assumed to interact by uniform linear forces
(`springs' whose rest length is the particle diameter).  Force chains are
evident.  Since non-cohesive grains cannot experience tensile forces, we have
repeated the simulation with a more realistic interaction in which `one-sided
springs' (which can  exert only compressional forces) have been employed. The
resulting force distribution is depicted in
Fig.~\ref{fig:lat2Dhexnotens_particles+forces}. Though in the second simulation
there is particle rearrangement (some contacts are severed, as observed in
\cite{Luding97} for a pile geometry) the appearance of the force chains in both
systems is very similar.  The force distribution vs. the horizontal coordinate,
at different depths, corresponding to Fig.~\ref{fig:lat2Dhex_particles+forces},
is presented in Fig.~\ref{fig:forcesdiffdepth}, and that corresponding to
Fig.~\ref{fig:lat2Dhexnotens_particles+forces} is presented in
Fig.~\ref{fig:notensionforcesdiffdepth}. These force distributions are
qualitatively similar though the latter
(Fig.~\ref{fig:notensionforcesdiffdepth}) is far closer to the experimental
findings~\cite{Geng00} on a similar system of disks (since it better represents
the properties of the grains).  Furthermore,
Fig.~\ref{fig:pertlowconlat2Dhex_particles+forces} presents the force chains
for a random network, obtained by adding a random number, uniformly distributed
in the range $[-\frac{d}{4},\frac{d}{4}]$, to the $x$ and $y$ coordinates of
points on a triangular lattice with lattice constant, $d$ (except those of the
bottom row), with springs connecting points whose distance is less than
$c_\mathrm{max}=1.2d$.  The results are again qualitatively similar to those
obtained in the experiment~\cite{Geng00}, the main difference with respect to
the case of a regular lattice being the fact that here the force chains are
somewhat shorter.

\begin{figure}[!ht]
\includegraphics[width=3.25in]{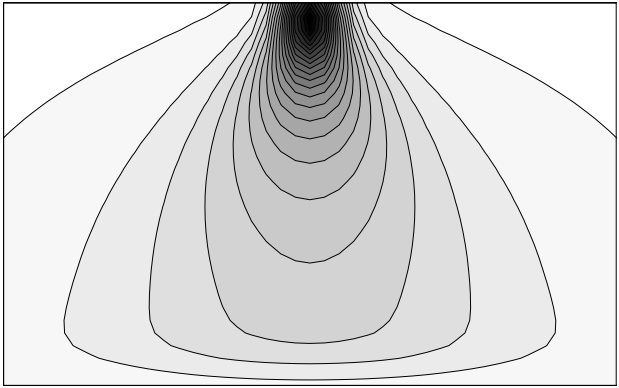}
\caption{Contour plot of $h\sigma_{zz}$, in the
  2D triangular lattice, in the region shown in
  Fig.~\ref{fig:lat2Dhex_particles+forces}. The contour spacing is
  $0.3$. Darker shades indicate larger values of  $|\sigma_{zz}|.$}
\label{fig:lat2Dhex_stress}
\end{figure}

Fig.~\ref{fig:lat2Dhex_stress} depicts the normal vertical stress component,
$\sigma_{zz}$, corresponding to the system in
Fig.~\ref{fig:lat2Dhex_particles+forces}, as calculated using the following
exact formula~\cite{Glasser01}:
\[
\sigma_{\alpha\beta}(\mathbf{r},t)= \frac{1}{2} \sum_{j\ne i} f_{i/j\alpha}\,
{r}_{ij\beta} \int_0^1 ds \phi(\mathbf{r}-\mathbf{r}_i(t) +s
\mathbf{r}_{ij}(t)),
\]
where $\mathbf{r}_i(t)$ is the center of particle $i$, $\mathbf{r}_{ij}(t)
\equiv \mathbf{r}_i(t) -\mathbf{r}_j(t)$, and $\mathbf{f}_{i/j}$ is the force
exerted on particle~$i$ by particle~$j$. Greek indices denote Cartesian
components.  The coarse graining function~\cite{Glasser01} is
$\phi(\mathbf{r})=\frac{1}{\pi w^2}e^{-(|\mathbf{r}|/w)^2}$, with \mbox{$w=d$},
the particle diameter, i.e., a fine resolution. The force chains are not
evident any more.  The model considered here corresponds, in the continuum
limit, to an isotropic 2D elastic medium~\cite{GoldenbergUP}.  Therefore one
can compare the stress obtained from the microscopic force distribution with
that of the corresponding elastic solution.

\begin{figure}[!ht]
  \includegraphics[width=3.25in]{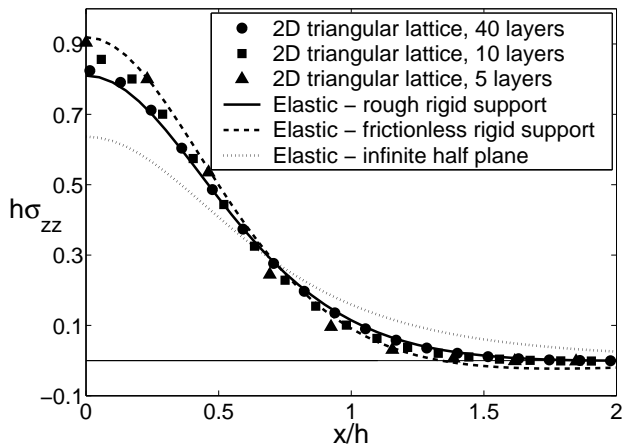}
\caption{$h\sigma_{zz}$ at the bottom of the 2D triangular lattice,
  compared to continuum elastic solutions. The applied force is unity. Note:
  $\sigma_{zz}\left(\frac{x}{h}\right)= \sigma_{zz}\left(-\frac{x}{h}\right)$.}
\label{fig:Lat2DHexCompareElastic}
\end{figure}

Fig.~\ref{fig:Lat2DHexCompareElastic} compares the vertical stress at the floor
of the system with elastic solutions for a finite slab (with rough or
frictionless support) and a half plane.  The convergence to the experimentally
appropriate~\cite{Clement} (rough support) solution for a sufficient number of
layers is evident. For the random system depicted in
Fig.~\ref{fig:pertlowconlat2Dhex_particles+forces}, the results are quite
similar, expect for (expected) fluctuations.

\begin{figure}[!ht]
  \includegraphics[width=3.25in]{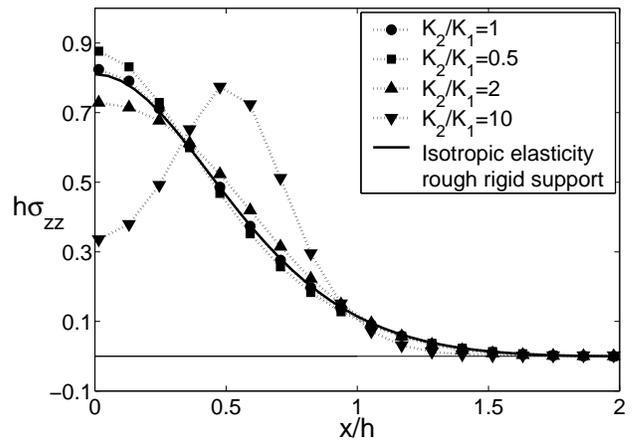}
\caption{$h\sigma_{zz}$ at the bottom of anisotropic triangular lattices
  composed of $40$ layers, compared to the isotropic elastic solution. The
  applied force is unity. Note: $\sigma_{zz}\left(\frac{x}{h}\right)=
  \sigma_{zz}\left(-\frac{x}{h}\right)$.}
 \label{fig:Lat2DHexAniso}
\end{figure}

\begin{figure}[!ht]
\includegraphics[width=3.25in]{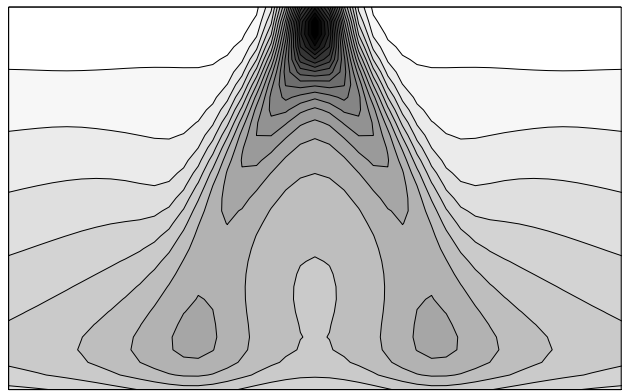}
\caption{Same as Fig.~\ref{fig:lat2Dhex_stress}, for the case of one-sided springs.}
\label{fig:lat2Dhexnotens_stress}
\end{figure}

Consider next an anisotropic medium, obtained by taking the spring constants
for contacts in the horizontal direction (parallel to the slab floor), $K_1$,
to be different from those for the contacts in the oblique direction, $K_2$.
As shown in Fig.~\ref{fig:Lat2DHexAniso}, the obtained stress distribution (on
the floor) is either single peaked (narrower than the isotropic one for
$K_2/K_1<1$, wider for $K_2/K_1>1$) or double peaked for sufficiently large
$K_2/K_1$. A similar double peaked distribution is obtained for the case of
`one-sided' springs, where some horizontal contacts are severed, corresponding
to the limit $K_2/K_1\rightarrow\infty$ for these contacts. This is evident in
Fig.~\ref{fig:lat2Dhexnotens_stress}, which clearly shows a macroscopic
anisotropy (compare to the macroscopically isotropic case shown in
Fig.~\ref{fig:lat2Dhex_stress}). These double peaked distributions are similar
to those obtained from hyperbolic models. The results presented here indicate
that phenomena similar to those suggested by the hyperbolic models can be
obtained using \textit{anisotropic} elasticity.  The equations of anisotropic
elasticity, which are of course elliptic, can approach hyperbolicity in the
limit of very large anisotropy (as mentioned  in~\cite{Cates99}).

\begin{figure}[!ht]
  \includegraphics[width=3.25in]{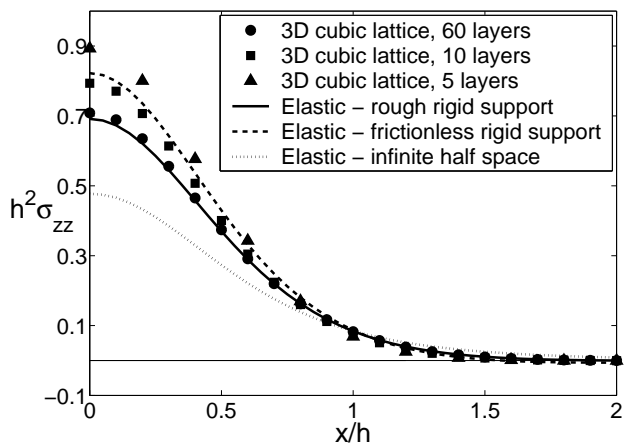}
\caption{$h^2\sigma_{zz}$ along the $x$-axis at  the bottom of a cubic
  lattice, compared to continuum elastic solutions. The results are scaled by
  $h$. Note:
  $h^2\sigma_{zz}\left(\frac{x}{h}\right)=h^2\sigma_{zz}\left(-\frac{x}{h}\right)$.}
\label{fig:Lat3DCompareElastic}
\end{figure}

\begin{figure}[!ht]
\includegraphics[width=3.25in]{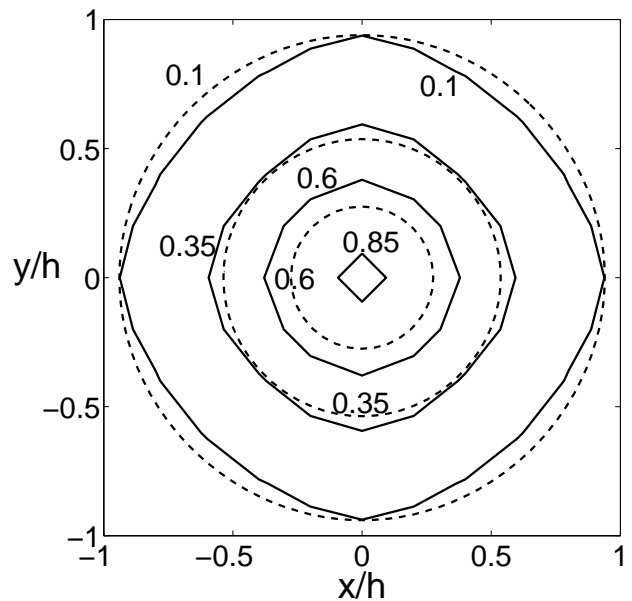}
\caption{A contour plot of $h^2\sigma_{zz}$ at the bottom of 3D slabs
  composed of 5 (solid lines) and 20 (dashed lines) discrete layers.  A unit
  force is applied at the top. The cubic anisotropy is evident for $5$ layers,
  while the distribution appears isotropic for $20$ layers.}
\label{fig:3Dbottomcontour}
\end{figure}

Very similar results to those presented above are obtained for a three
dimensional (3D) system: consider a collection of particles positioned on a
simple cubic lattice.  As shown in~\cite{Kittel56}, this system corresponds, on
large scales, to an isotropic elastic continuum when the spring constants for
springs coupling nearest neighbors equal those coupling next nearest neighbors.
While this model clearly does not describe the interaction of granular
particles, we believe it is useful for the description of the crossover between
microelasticity and macroelasticity.  Furthermore, as shown below, the stresses
computed for this model are in close correspondence with the experimental
results reported in~\cite{Clement}.  Consider a 3D slab of finite height, with
a downward unit force acting on a particle in the top layer.  As in the 2D case
described above, the results for discrete lattices converge to the appropriate
finite slab elastic solution with increasing depth
(Fig.~\ref{fig:Lat3DCompareElastic}).  The effect of small scale anisotropy
(here, the cubic symmetry) is seen in Fig.~\ref{fig:3Dbottomcontour}, which
depicts contour plots of the distribution of the vertical stress on the floor.
The underlying cubic symmetry is evident for a small depth of the system and is
washed out for larger systems, where continuum elasticity holds.

Our results for a shallow discrete slab are similar to those obtained
experimentally for a `loose' granular packing~\cite{Clement}, for which the
stress distribution is narrower than that predicted by continuum elasticity.
In contrast, for a dense packing, experiments show a wider distribution. As
mentioned in~\cite{Clement} and above, this can be explained by anisotropic
effects (recall that the above anisotropic 2D system can exhibit both narrower
and wider distributions).  The method of preparation of the experimental
system~\cite{Clement} suggests that a relevant model may consist of a number of
isotropic elastic layers of variable moduli. Indeed, in this case the stress
distribution on the floor can be wider or narrower than the solution for a
homogeneous slab, depending on the distribution of moduli of the
layers~\cite{GoldenbergUP,Savage97}. In addition, the stress distribution
becomes wider the smaller the rigidity of the support
\cite{GoldenbergUP,Savage97}.

\begin{acknowledgments}
  Support from the Israel Science Foundation, grants no. 39/98 and
53/01, is gratefully
  acknowledged.
\end{acknowledgments}


\begin{thebibliography}{13}

\bibitem{ChaosGranularFocus} Focus Issue on Granular Materials, \textit{Chaos}
  \textbf{9} (1999).
  
\bibitem{Nano} e.g., J.-P. Salvetat et~al., Appl. Phys. A \textbf{69},
  255 (1999); J. Broughton et al., Phys. Rev. B \textbf{56}, 611 (1997).

\bibitem{Cates99} M.~E. Cates, J.~P. Wittmer, J.-P. Bouchaud and P. Claudin,
  Chaos \textbf{9}, 511 (1999) and refs. therein.
  
\bibitem{ForceChains} S.~N. Coppersmith et~al., Phys. Rev. E \textbf{53}, 4673
  (1996); F. Radjai, S. Roux and J.~J. Moreau, Chaos \textbf{9}, 544 (1999);
  J.-P. Bouchaud, P. Claudin, D. Levine and M. Otto, Eur. Phys. J. E
  \textbf{4}, 451 (2001).

\bibitem{Nedderman92} R.~M. Nedderman, \textit{Statics and Kinematics of
    Granular Materials} (Cambridge University Press, 1992).

\bibitem{Savage97} S.~B. Savage, in \textit{Powders and Grains 97}, 185, Ed.
  R.~P.  Behringer and J.~T. Jenkins (Balkema, 1997).
  
\bibitem{Geng00} J. Geng et~al., Phys. Rev. Lett.  \textbf{87}, 035506 (2000).
  
\bibitem{Clement} G. Reydellet and E. Cl{\'e}ment, Phys. Rev. Lett.
  \textbf{86}, 3308 (2001); D. Serero et~al., Eur. Phys. J. E \textbf{6}, 169
  (2001).

\bibitem{Luding97} S.~Luding, Physical Review E {\bf 55}, 4720 (1997). 

\bibitem{Glasser01} B.~J. Glasser and I. Goldhirsch, Phys. Fluids \textbf{13},
  407 (2001).

\bibitem{GoldenbergUP} C. Goldenberg and I. Goldhirsch, unpublished.
  
\bibitem{Kittel56} C. Kittel, \textit{Introduction to Solid State Physics
    (Second Edition)} (Wiley, 1956).
  
\end{thebibliography}
\end{document}